# Evaluating the Usability of Automatically Generated Captions for People who are Deaf or Hard of Hearing

Sushant Kafle
Rochester Institute of Technology
152 Lomb Memorial Drive
Rochester, NY, 14623
+1-585-475-2700
sushant@mail.rit.edu

Matt Huenerfauth
Rochester Institute of Technology
152 Lomb Memorial Drive
Rochester, NY, 14623
+1-585-475-2700
matt.huenerfauth@rit.edu

## ABSTRACT

The accuracy of Automated Speech Recognition (ASR) technology has improved, but it is still imperfect in many settings. Researchers who evaluate ASR performance often focus on improving the Word Error Rate (WER) metric, but WER has been found to have little correlation with human-subject performance on many applications. We propose a new captioning-focused evaluation metric that better predicts the impact of ASR recognition errors on the usability of automatically generated captions for people who are Deaf or Hard of Hearing (DHH). Through a user study with 30 DHH users, we compared our new metric with the traditional WER metric on a caption usability evaluation task. In a side-by-side comparison of pairs of ASR text output (with identical WER), the texts preferred by our new metric were preferred by DHH participants. Further, our metric had significantly higher correlation with DHH participants' subjective scores on the usability of a caption, as compared to the correlation between WER metric and participant subjective scores. This new metric could be used to select ASR systems for captioning applications, and it may be a better metric for ASR researchers to consider when optimizing ASR systems.

## Keywords
Accessibility for People who are Deaf or Hard-of-Hearing; Automatic Speech Recognition; Real-time Captioning System; Caption Usability Evaluation

## 1. INTRODUCTION
Over 37 million people in the U.S are Deaf and Hard-of-Hearing (DHH) [4], and many benefit from offline captioning (e.g., for pre-recorded television programming) or real-time captioning services (e.g., in classrooms, meetings, and live events). Over the past few decades, automated speech recognition (ASR) technologies have seen major progress in their accuracy and speed. With its increasing maturity, ASR technologies are now being used commercially for many consumer applications. Due to its low cost and scalability, ASR has potential for the task of captioning [2][11][13]. Broadly, our research investigates the use of automatic speech recognition (ASR) to provide captioning services for DHH users, especially in real-time contexts such as meetings with hearing colleagues. The focus of this paper is on designing and evaluating a metric for evaluating the quality of an ASR system, to determine whether its output would produce understandable captions for DHH users.

Section 1 describes the use of ASR for captioning and motivates why the most common evaluation metric for ASR systems (Word Error Rate, WER) is insufficient for evaluating ASR for generating captions for DHH users. Section 1.3 introduces our new metric and outlines our major research questions. Section 2 describes prior work on the limitations of WER and on identifying new evaluation metrics. Section 3 provides details about the design and development of our new metric. Section 4 outlines the hypotheses examined in this paper. Section 5 describes the stimuli preparation strategy and the methodology of our study with DHH participants. Finally, sections 6 and 7 present our results and conclusions.

### 1.1 Automated Captioning Technologies
Real-time captioning services are usually provided by a human transcriptionist who transcribes the audio information to text using a keyboard, with the captions being displayed on a screen. Although well-trained transcriptionists can produce accurate real-time captions with a speed of over 200 words per minute, systems that rely on trained transcriptionists, e.g. Computer-Aided Access Real-Time (CART) or similar services [39], are not suitable for impromptu meetings or extremely brief conversational interactions, given the overhead cost of arranging a transcriptionist. Recently, a crowd-sourced technique to generate real-time captions [26] has been shown to produce accuracy comparable to that of a trained transcriptionist, with lower cost. However, the proposed system relies on the Amazon Mechanical Turk platform, and the business model for providing captioning services is not yet clear.

Researchers have also investigated architectures for providing real-time captioning using ASR, including: use of an ASR system to generate automated captions directly [2][11][13] or the use of non-experts to inspect the output of an ASR system [20][40][41]. While asking humans to correct errors in captions generated by ASR sounds promising, it has been found that manually identifying and correcting errors in ASR generated captions can take more time than producing the transcript without the support of ASR [16]. With recent improvements in ASR, recent work has investigated the potential of ASR technologies (without any human intervention) for captioning live meetings [9] or for classroom lectures [23].

### 1.2 Word Error Rate as an ASR Metric
Accurate, large-vocabulary, continuous speech recognition is still considered an unsolved problem. Although there have been recent leaps in the performance of these systems [43], ASR performance is generally not on par with humans, who currently provide most caption text for DHH users. Noise in the input audio, the ambiguity of human speech, or unforeseen speaker characteristics (e.g. a strong accent) can lead to ASR errors. As researchers continue to improve ASR accuracy, they generally report the performance of their systems using a metric called Word Error Rate (WER). Given the ubiquity of this metric, it is reasonable that reducing WER may be a goal of many ASR research efforts (implicitly, if not overtly).

$$WER = \frac{S + D + I}{N}$$

**Figure 1: Formula for Word Error Rate (WER), based on *S* (number of erroneous substitutions of one word for another), *D* (number of deletions, i.e. erroneous omissions of words that were spoken), *I* (number of insertions of spurious words in the ASR output), and *N* (number of words actually spoken).**

As shown in Figure 1, WER is calculated by comparing the "hypothesis text" (the output of the ASR system) to the "reference text" (what the human actually said in the audio recording). The metric considers the number of misrecognitions in the hypothesis text, normalized by the word-length of the reference text. Notably, WER does not consider whether some words may be more important to the meaning of the message or whether some words might be more predictable than others in a text. In fact, researchers have previously found that humans perceive different ASR errors as having different degrees of impact on a text – some errors might distort the meaning of the text more harshly than others [31]. Others have found that the impact of errors may be dependent upon the specific application in which ASR is used [10][33].

### 1.3 Metric of ASR Quality for DHH Readers

The premise of this paper is that rather than simply counting the number of errors, it would be better to consider which words are incorrect or where they occur in the sentence, when evaluating ASR text output for captioning applications for DHH users. As discussed in section 2, some researchers have previously examined the limits of the WER metric and have considered some alternatives. A novel aspect of our research is that we are specifically interested in measuring the quality of ASR output for a captioning application for DHH individuals, and we evaluate our proposed metric in a user-study with DHH participants.

There are reasons to believe that it is important to create and evaluate metrics of ASR output quality specifically targeted for DHH users. Anecdotally, some accessibility researchers have argued that ASR-generated errors on captions are more comprehension-demanding than human produced errors [2][25]. Further, there are known differences in literacy rates and reading mechanisms among DHH readers, which differ from their hearing peers: In standardized testing in the U.S., English literacy rates have been measured to be lower for deaf adults [21][29]. Furthermore, literacy researchers have hypothesized that the basic mechanism employed by many deaf adults to understand written sentences differs from hearing readers: That is, deaf readers may identify the most frequent content words and derive a complete representation of the meaning of the sentence, ignoring other words [3][6]. This reading strategy is often referred to as a "keyword" strategy, and it suggests that a subset of the words in a caption text might be of very high importance to DHH users (for text understandability). Following this same reasoning, it might be disadvantageous to penalize each error in a caption text equally. Some errors may be very consequential to the understandability of the text (with the potential to mislead or confuse the readers), while other errors may have little impact (perhaps easily ignored by readers). Our goal is to develop a metric that can predict the quality of an ASR text output based on the usability of the text as a caption for DHH users. Unlike WER, we want our metric to distinguish between harmful errors in the caption (likely to degrade the quality of caption for DHH users) and less harmful errors; the metric should use this distinction when penalizing a text for each type of error.

With this aim, we developed the Automated-Caption Evaluation (ACE) metric, which measures the impact of errors on the understandability of a caption for DHH users. ACE makes use of a word predictability score (entropy of a word given its context) as a measure to identify keywords in a text and, it uses semantic distance (between the error words and actual words in the reference text) as an approximation of the deviation in meaning due to errors. Using these two measures, ACE computes an impact score for the errors in the text, which is used to predict the degree of usability of the ASR output as a caption text for DHH users.

The focus of this paper is on comparing the performance of our ACE metric on this task and to explore two research questions:

Q1. Do DHH users subjectively rate ASR-generated captions preferred by the ACE metric as having greater usability than ASR-generated captions that are not preferred by ACE?

Q2. Does the ACE metric correlate better with the subjective quality judgments of DHH users on ASR-generated captions, as compared to their correlation with WER scores?

## 2. PRIOR ASR EVALUATION RESEARCH

This section surveys prior research on the limitations of WER as an evaluation metric for ASR research, efforts to design alternative metrics, and methods to identify important words in a text for DHH readers – which is an aspect of our new alternate evaluation metric.

### 2.1 Limitations of WER

While WER has been the most commonly used evaluation measure for speech recognition performance, spoken language technology researchers have argued for alternative evaluation measures that would better predict human performance on tasks that depend on ASR text output usability [30][32]. While WER has a lower bound of zero (indicating that a hypothesis text is a perfect match for a reference text), researchers have criticized WER since it lacks an upper bound [30], making it difficult to evaluate WER scores in an absolute manner. Further, some researchers have argued that WER is ideally suited to evaluation of ASR quality only for those applications in which the human can correct errors by typing, since the WER metric is based upon counting errors – which directly relates to the cost of restoring the output word sequence to the original input sequence [30].

In other applications, researchers have observed a weak relationship between WER and human task performance. For example, in the task of spoken document retrieval (in which a user searches for a speech audio file, that has been transcribed by ASR, by typing search terms for desired information), researchers have found that the WER of the ASR system has little correlation with the retrieval system performance [15][18]. Moreover, in [42] researchers saw improvements in a spoken language understanding task, even during a significant increase in WER.

### 2.2 ASR Evaluation Methods

Several researchers have proposed alternative metrics to WER for evaluating the performance of ASR for specific applications:

Some researchers have weighted errors based on their Term Frequency-Inverse Document Frequency (TF-IDF) measure [33]. TF-IDF is commonly used by researchers studying information retrieval; it assigns high scores to words that are generally "rare" but which appear in great frequency in a particular document, e.g., if a rare word like "penguin" appears very frequently on a particular webpage, then it is reasonable to think that the word "penguin" is an important "keyword" for that webpage. Researchers have used TF-IDF as a "loss function" during the decoding step of their ASR system [33]. (During decoding, the ASR system aims to determine the most likely sequence of words that corresponds to speech information.) The loss function penalized errors on keywords more heavily than errors on other words, when choosing from a list of output candidates. The authors explored using this metric as a weighting factor in a Boolean fashion (keyword or non-keyword) or by using the actual numerical TF-IDF scores as weights.

Researchers in [15] attempted to modify WER to weight "content words" more heavily than other words. Generally speaking, content words include nouns, verbs, adjectives, and adverbs that convey semantic meaning, rather than "function words," e.g. determiners, that convey grammatical information. The authors used ASR in an information retrieval application; users searched for content in spoken audio recordings. The authors found a nearly linear relationship between the proposed metric and retrieval performance across different systems: i.e., ASR systems that recognized content words more accurately provided the best input for their retrieval task. Thus, both [15] and [33] found that keyword identification (to differentially weight specific kinds of errors) led to useful ASR metrics for applications related to information search.

Some researchers have considered applications of ASR that are even closer to our focus on automatic captioning: For instance, some have proposed a metric for evaluating ASR output on a speech transcription task [31]; their metric was based on opinion scores collected from humans who judged the quality of ASR-generated voicemail-to-text transcripts. Scores from their metric correlated with the human judgments better than WER did. Their metric learned the cost of different error types (namely, insertion, deletion, and substitution) and learned a weight factor called the "saliency index" to predict the text understandability. While not focused on creating a fully automatic metric, researchers in [1] investigated different categories of captioning errors (e.g., substitution of a word with an incorrect tense) and weighted each category to design a "weighted WER" metric. This metric was proposed for evaluating the accuracy of captions for television.

The "match error rate" (MER) and "word information loss" (WIL) metrics were introduced in [32], as replacements for WER in settings where high error rates are common. The MER metric is similar to WER except that it is properly normalized and thus computes the "probability" of a given match (between the reference text and the hypothesis text) being incorrect. Similar to MER, WIL is a probabilistic approach that approximates the proportion of the word information lost due to the presence of errors.

Our work is also heavily inspired by the work of [30], which discusses the problems of application-oriented evaluation of ASR systems. In their work, the researchers proposed a generic framework to evaluate the ASR output based on information retrieval concepts like "precision" and "recall." Their proposed framework assumes that the speech recognition task is analogous to an information retrieval task, i.e. the goal for transcription is to retrieve all the relevant information (i.e. the spoken word) from in the original speech signal. Their framework also provided room to incorporate application-dependent importance weights for words and for different ASR error types. However, for our application of real-time captioning for DHH users, the assumptions made in their framework are less appropriate: they treated words as independent units of information, without considering their position in a sentence, i.e. under this assumption identical words located at different positions in a sentence will have identical weights.

In contrast, section 3 describes how our new metric considers the predictability a word at its position in a sentence. Motivation for considering word-predictability in context is discussed in section 2.3. In addition, our new metric considers the semantic deviation between the error word and the actual reference word (section 3).

## 2.3 Word Predictability

We previously discussed how some researchers [6][7] have hypothesized that deaf readers use a sentence-understanding strategy in which they seek content words in order to derive a representation of sentence meaning, potentially ignoring other information, e.g. morphosyntactic relationships between words. Research on the eye movements of deaf readers has also revealed that deaf readers visually fixate on approximately 30% of the words in a text. The skipped words were largely determined by lexical factors such as how frequent a word is, the length of the word, and the predictability of the word in that sentence [3]. Similarly, in [37], researchers found that both the length of the word and predictability of the word in context were related to whether readers skip over a word and also on the amount of time readers spent on non-skipped words. In general, highly predictable words have been shown to be read faster and skipped more often than unpredictable words by most readers [34], even more so with less skilled readers [3].

Furthermore, word predictability has been a common theme in prior research on assessing the readability of a text or the reading comprehension skills of a participant [8][24][35][36] . For instance, the "Cloze procedure" is an assessment methodology that has been around for years and is one of the most common ways of evaluating both the readability of a text and the reading skills of participants. In this task, the participant is given a text with one word omitted, and they must guess the missing word. Most standardized English-language tests (e.g., TOEFL, GRE, WRAT) utilize some variation of the Cloze technique to evaluate participants' reading skills.

The predictability of a word refers to the degree to which a reader can use the context to guess the word. For example:

*The _____ was barking at the mail-man.*

The predictability of the word *dog* is high given the context. The context of the word is powerful enough to provide a hint as to what the word is. Conversely, in the sentence:

*The meeting is scheduled on _____.*

The predictability is very low – suggesting that the readers might not be able to rely on the context to predict the word.

Given the use of word predictability in reading assessment (Cloze tests) and given the aforementioned eye-tracking research (indicating that DHH readers are more likely to skip over highly predictable words), we decided to include word predictability in our new measure of ASR output quality for captioning for DHH users. Specifically, section 3 presents how we estimate word predictability and use it to identify important keywords in the sentence – ultimately allowing us to invent a new metric for evaluating the impact of ASR errors on caption usability.

|  | **Caption Text** | **WER** | **ACE** |
|---|---|---|---|
| **Reference Text** | based on the information we gather we will send it off to the lead recruiter for each of those teams | -- | -- |
| **ASR Hypothesis 1** | on the information we gather we will send it off to *relief worker* for each of those *chains* | 0.25 | 0.65 |
| **ASR Hypothesis 2** | based the information gather will send it off the lead recruiter for each those teams | 0.25 | 0.28 |

Table 1: This example demonstrates how ACE penalizes captions containing different errors, as compared to how the WER metric evaluates each. Higher ACE or WER values indicate "worse" output for each metric.

## 3. AUTOMATED-CAPTION EVALUATION

As discussed in section 1, we are interested in the potential for ASR systems to be used as a real-time captioning tool for impromptu meetings. There are many commercial and research ASR systems available, each with different capabilities, e.g., adapting to the voice of specific speakers, operating in contexts with different types of background noise, or recognizing different vocabulary or genres [19][27][28]. A natural question is how to compare ASR systems to determine their suitability for use in this context, and given the limitations of WER discussed in section 2, we therefore present a new Automated-Caption Evaluation (ACE) metric.

ACE considers two primary factors: (a) the predictability of a word given its context and (b) the semantic deviation between the error word and the reference word. These two factors are used to predict the impact of an error in the caption text as follows:

$$I(e_i) = \alpha * E(w_i) + (1 - \alpha) * D(w_i, e_i) \quad (1)$$

where $E(w)$ represents the predictability score of the reference word $w$, $D(w, e)$ represents the semantic distance between the reference word $w$ and the error word $e$ and $I(e)$ represents the impact score of the error $e$. Alpha ($\alpha$) represents the interpolation weight, which determines how much each factor contributes to the overall impact score. In other words, the overall impact of an error is determined by the weighted combination of the predictability of the word and the semantic distance of the error and the reference word. The weighting factor is determined by the value of alpha ($\alpha$). Section 3.4 describes how this per-word impact score $I(e)$ is used to calculate an overall ACE metric for an entire sentence.

To preview for the reader the benefits of this new metric (before beginning our detailed discussion of the technical details behind its calculation), we provide Table 1, which demonstrates how ACE evaluates caption quality, as compared to WER. Two ASR output texts are shown that have identical WER scores; however, our ACE metric assigns different scores to the texts. ASR Hypothesis 1 has errors on words that are less predictable and the semantic difference between the substituted words and their original reference is great (e.g. "chains" vs. "teams"); for this reason, the ACE metric assigns a higher score (indicating worse output) to that text.

The following sections explain how we calculate key components of this metric and how we use them to create our final metric. An evaluation of the efficacy of this metric (in a study with DHH participants evaluating captions) is presented in section 3.3.

### 3.1 Word Predictability Sub-Score

To compute the predictability score of a word, we utilized several n-gram language models; these models are based on how frequently certain sequences of words, of various length, have appeared in large collections of text. Similar models are commonly employed in word-prediction systems for text-entry applications, e.g. [14]. Based on the probability score assigned to the predictions by the language model, we compute the predictability score for the word.

#### 3.1.1 Language Model

We trained our n-gram models (n = 1 to 5) on Switchboard [17], English CALLHOME and TED-LIUM [38] corpora, which contain a total of 1.9 million unique words. These corpora were selected because they closely represent text from conversational speech dialogs (similar to the one-on-one meeting context in which we are considering the use of ASR for real-time captioning).

The n-gram models were used bidirectionally, to make predictions using both left and right word-sequence contexts, independently. To rank the possible word candidates using each context, a Stupid Back-off [5] mechanism was utilized. For ranking predictions from left context, the following scoring function was used:

$$S(w_i | w_{i-k+1}^{i-1}) = \begin{cases} \dfrac{\text{count}(w_{i-k+1}^i)}{\text{count}(w_{i-k+1}^{i-1})} & \text{if count}(w_{i-k+1}^i) > 0 \\ \lambda\, S(w_i | w_{i-k+2}^{i-1}) & \text{otherwise} \end{cases}$$

where a recommended value of 0.4 was used for lambda ($\lambda$) [5]. A similar scoring function was used to rank the candidates from the right context. The predictions from both the right and left contexts were combined and then ranked for later use.

#### 3.1.2 Word Entropy

To obtain a predictability score from these predictions, we first selected the top ($N_c$=20) ranked unique candidates and transformed their count probabilities to normalized probabilities (that sum to 1). For instance, in the "the meeting is scheduled on ___" example, the language model might predict various possible words, e.g. "Monday, Friday, Tuesday, etc." The model will predict that each of these words has some probability of appearing in that context.

Based on the distribution of probability among the candidates, an entropy score was calculated as follows:

$$E(w) = \sum_{i=0}^{N_c} -P(w_c(i)) * \log(P(w_c(i)) \quad (2)$$

where $E(w)$ represents the entropy of a word $w$ (at a unique position in the text). $w_c(i)$ is a candidate of the word $w$ predicted by the language model and $P(w_c)$ is the probability of the candidate $w_c$ as determined by the language model.

The entropy score calculated in Equation 2 is a measure often used in information theory to calculate the unpredictability of a state. In our application, it is the measure of the degree of unpredictability of a word given the context. A higher value indicates that the chances of picking the right word from the list of candidates is low – meaning it is difficult to predict the word. Whereas, a lower value would indicate that some words in the list of candidates clearly have higher probability than others – meaning it is easier to predict the word. The entropy is normalized to get a predictability score within a (0, 1) range. Figure 2(a) provides a visual example of how this metric assigns scores to a text. As shown in the figure, some words in the example text are highly unpredictable (with higher entropy score) – like *requested, college,* etc., while some are fairly predictable – like *is, to,* etc. in the text.

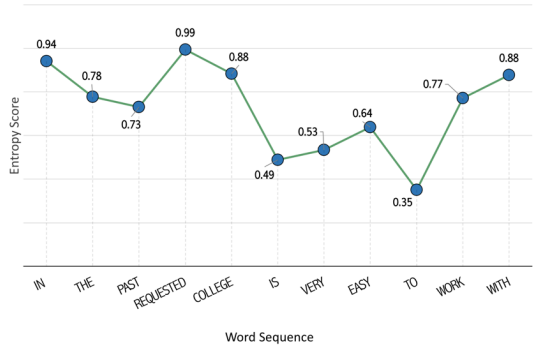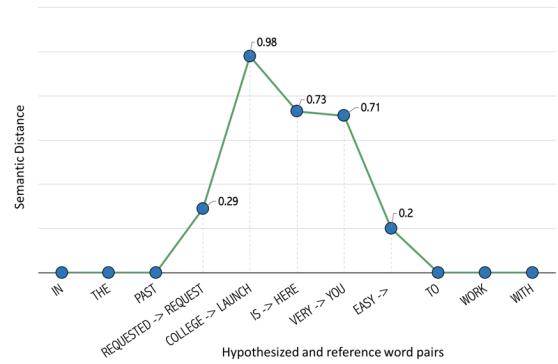

(a) Word Predictability Score    (b) Semantic Distance

Figure 2: Word predictability and semantic distance values are demonstrated here for an example sentence. (a) Each word is assigned an entropy score where the greater score indicates the unpredictability of the word. (b) Each hypothesized word is aligned with the reference word (reference word -> hypothesized word), and for each error, semantic distance is approximated. Greater distance score indicates larger deviation in the spoken message due to the error word.

In conclusion, for each error word, the entropy score of the corresponding reference word ($w$) is calculated to estimate the word predictability score ($E(w)$). For insertion errors, where there could be two adjacent reference words that could be responsible for the error, an average entropy score is computed based on the adjacent reference words.

### 3.2 Semantic Distance Sub-Score

The second factor that is considered in our ACE metric is the degree to which the meaning of a word in the output text differs from the meaning of the actual word that was spoken. In order to compute this semantic disagreement between the error word and the actual reference word, we utilized a pre-trained word2vec tool from Google[1]. The Word2vec tool provides a vector representation of words which can be subsequently used in many natural language processing applications and research. In this paper, the word2vec tool is used to compute semantic distance between the two words.

The semantic distance measure is used to represent the cost of an error. As shown in Figure 2(b), semantic distance score for minor errors (for e.g., *'requested' -> 'request')* would receive a small score, while an extreme misrecognition (for e.g., *'college' -> 'launch'*) would receive a high semantic distance score. However, for insertion and deletion errors, the length of the word is used to approximate the distance. For insertion and deletion errors, a constant value of 0.05 was chosen to scale the word length to get the semantic distance which was decided empirically from the analysis of errors from ASR system on a separate dataset[2].

### 3.3 Selecting a Weight between Sub-Scores

The word predictability sub-score (section 3.1) and the semantic distance sub-score (section 3.2) are combined using a weighted sum to produce an error impact score (as shown in Equation 1), but that equation requires us to select a tuning parameter alpha ($\alpha$) to specify how much each sub-score contributed to the overall error impact score. To learn the appropriate value of alpha ($\alpha$), we "fit" the value of this parameter, using response data provided to us by researchers who had conducted a prior study with DHH participants regarding the impact of various ASR errors in captioning [22], with the following demographic characteristics: ages 20 to 32 (mean = 22.63 and deviation= 2.63), 12 men and 18 women, 26 participants self-identified as Deaf and 4 as Hard-of-Hearing.

In that prior study, DHH users were presented with imperfect English text passages (containing errors that had been produced by an ASR system when the text had been spoken aloud and then automatically recognized), and they were asked to answer questions that required understanding the information content of those passages. Each question was based on information from a single sentence in the passage, and each sentence contained 0 or 1 ASR errors. Participants received a score of 1 if they answered a question correctly, and 0, if answered incorrectly.

From all the data collected during that prior study, we examined the subset of question-responses that corresponded to English sentences that had contained ASR errors. This data was used to calculate a "comprehension score" for each sentence, by averaging the scores from the 30 participants for questions about that sentence. For each sentence, we examined the ASR error within and calculated word-predictability and semantic-distance sub-scores. Given this new dataset, the value of the alpha parameter was tuned using a grid search to maximize the prediction accuracy of our impact score metric in Equation (1). We obtained an alpha ($\alpha$) value of 0.65, suggesting that in our task the impact score depended slightly more on the word predictability score, rather than on semantic distance.

### 3.4 Metric Formulation

The ACE metric makes use of the impact scores of the errors in the text to compute the final evaluation score. One way to formulate our metric would have been to compute this score is by summing over the impact scores of each error and normalizing it by the length of the reference text, as had been done for WER in Figure 1, e.g.

---

[1] https://code.google.com/archive/p/word2vec

[2] https://catalog.ldc.upenn.edu/ldc93s1

$$\text{Version1\_ACE} = \frac{\sum I(e_i)}{N}$$

where $I(e)$ is the impact of error ($e$) on the usability of the captions and, $N$ is the length of the reference text. This style of formulation for a metric is quite common [1][31][32], and its value would be equal to WER if all the error impact scores were to equate to 1. However, this above formulation of the metric has some limitations:

- For each error ($e$) in the hypothesized text, we compute an impact score by considering a single error ($e$) at a time in the reference text – consequently ignoring any non-linear impact due to a cascade of errors (effect of an error in presence of others). Thus, a sum of impact scores of individual errors is insufficient for capturing the true impact of such situations.
- Since the impact score of each error lies between the 0 and 1, the metric will always be less than or equal to WER, i.e.

$$0 \leq \text{Version1\_ACE} \leq WER$$

This is because, $N$ (the length of the reference text) is not the true normalizing constant for our metric.

Due to these limitations above, we invented a new (and final version) of the ACE metric as follows:

$$\text{ACE} = \frac{\max_{e \in E} I(e)}{\log(N) - \log(n)}$$

where E is the set of all the errors (substitution, insertion and deletion) in the hypothesized text. $I(e)$ represents the impact of each error, defined in Equation (1). $N$ is the length of the reference text, and $n$ is the total number of errors in the hypothesized text.

In this new version, the impact due to errors the *maximum* score among all the individual impact scores $I(e)$ for all errors in the text. When calculating alternative versions of this metric on example texts, we found that taking the *maximum* value represented our subjective impression of error impact across a single sentence better than other common alternatives, e.g. mean or median. Intuitively, if a sentence contains a major error, the overall effect on the sentence understandability can also be major; furthermore, the number of additional words in the text does not subjectively "improve" the sentence in a linear manner, which would be suggested by our Version1_ACE metric shown previously. There are limitations of this approach, e.g. extending this approach to longer, multi-sentential texts; section 7 discusses some limitations.

Similarly, the final version of our metric discounts the effect of $N$ on the impact score, as compared to the previous version. As the length of the reference text increases, it slowly mitigates the impact of individual errors – the rationale being that as readers have more context (more surrounding words), it is easier to decipher the true meaning of the text. The rate of this change is regulated in a non-linear fashion using log. But, if the number of error increases with the reference text, the impact of errors is counterbalanced (note the subtraction of a log(n) term in the denominator).

Like WER, ACE is also an error measure, meaning that a *lower* ACE score indicates a *better* caption text. Similar to WER, ACE does not have an upper bound; however, we note that it is trivially possible to modify the metric to prevent it from exceeding some limit, e.g. $\max(1, \text{ACE}(W))$ would prevent it from exceeding 1.

In summary, our intention when designing this new ACE metric was to penalize ASR output texts that contain errors that are likely to lead to misunderstanding; specifically, the ACE metric considers errors at locations in a text that are less predictable and errors that deviate semantically from the actual word.

## 4. EVALUATING THE EFFICACY OF ACE

While the goal of our overall project is to develop a metric that could automatically evaluate the usability of a caption for DHH users, we must select a more specific scope in order to design a study to evaluate the efficacy of this new metric. We have selected to focus on measuring the efficacy of ASR for providing captions during a business meeting between a hearing person (speaking English) and a DHH participant. The rationale for this focus is that impromptu one-on-one meetings in a workplace may be a situation in which it is unlikely for professional captioning services to be available. Therefore, there is an opportunity for using automatic methods like ASR technology as a captioning tool. With this focus, we predict that DHH users will subjectively prefer ASR text output that is predicted as less erroneous by our ACE metric (as compared to ASR text output predicted as being less erroneous by the traditional WER metric).

Specifically, if we ask DHH users to evaluate the quality of ASR text output, we hypothesize the following:

**H1**: If we compare ASR output texts predicted as better by WER (i.e. with low WER-to-ACE ratio) to ASR output predicted as better by ACE (i.e. with high WER-to-ACE ratio), DHH participants will subjectively prefer texts preferred by ACE.

**H2a**: The subjective preference judgments of DHH participants on ASR output texts will correlate significantly with ACE.

**H2b**: There will be a significantly higher correlation between DHH human judgments and ACE, as compared to the correlation between DHH human judgments and WER.

## 5. OVERVIEW OF STUDY DESIGN

To investigate these hypotheses, a comparative evaluation study was designed where the performance of the evaluation metrics (ACE and WER) was assessed based on their agreement with DHH participant's judgment scores on the usability of captions generated by an ASR system.

### 5.1 Preparation of Stimuli

To create caption texts to display to our DHH participants during the study, we made use of some staged and prerecorded videos (provided to us by Christopher Caulfield, details in Acknowledgements). These videos display one side of a two-person business meeting communication – the speaker leading the conversation in the video is made to look like he is interacting to the participant who is watching the video, as shown in Figure 3.

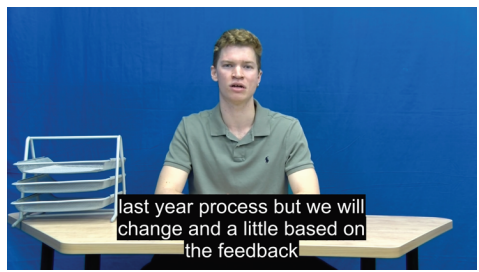

**Figure 3: Preparation of a fake meeting transcript.**

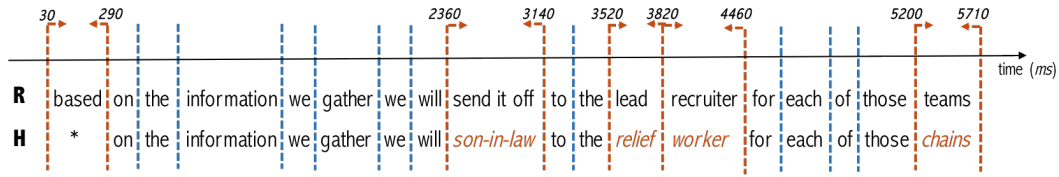

**Figure 4:** Time based alignment of reference (R) and hypothesized (H) text. The grouping with red dotted arrowhead lines indicates individualized errors aligned with corresponding reference text based on word level timestamps.

We extracted the verbatim script of what the human actor said during the videos, and we used the entire text of this video as a potential source of stimuli sentences for inclusion in this study (details in 5.1.1 and 5.1.2 below). Next, we processed the original audio recording from these videos using an ASR system that we expected to make a large number of errors (it is important for our stimuli selection process described in 5.1.2 for us to have many possible errors to choose from). For this processing, we used the CMU Sphinx 4 system with its off-the-shelf US English acoustic and language models which have been previously disseminated to the research community[3].

While a simplistic approach for creating stimuli for the study would have been to simply display the raw output of the ASR system to users, we were interested in obtaining judgments from participants on texts that had a variety of ACE metric scores. Furthermore, to investigate hypothesis H1, we were interested in presenting users with some pairs of ASR text output that displayed multiple hypotheses (i.e. two different guesses from the ASR system about what it heard), with one of the texts having a low WER-to-ACE score ratio (indicating that WER believed the text to be good, but ACE did not) and the other with a high WER-to-ACE ratio. Since ASR systems actually consider a wide variety of hypotheses when they analyze a speech audio file (with one hypothesis correct, and the remainder containing some variety of errors), we wanted to search the space of ASR output candidate hypotheses to select texts to display in our study with various WER-to-ACE ratios. Sections 5.1.1 and 5.1.2 describe our procedure for identifying ASR output hypotheses to display in our study with diverse WER-to-ACE ratios. Rather than inventing artificial errors to insert into the texts, our procedure obtains a large number of real ASR errors on a text and selects a subset of these errors to include in the texts displayed.

### 5.1.1 Time-based Alignment

After we prepared the meeting script and ran it against our low-accuracy ASR system, the next step was to align the reference text (the verbatim script of what the human actually said) and the hypothesis text (the output of the ASR system) to obtain a list of all the errors in the ASR output. While the ASR output hypothesis text already included timestamps of when the ASR believed each word had been spoken, we needed to identify timestamps for each word in the reference text. We manually compared the reference text to the original audio to obtain timestamp values for each word.

Next, we needed to time-align the hypothesis text to the reference text, to correctly identify all the errors in the hypothesis text. Standard alignment tools like [12] were ill-suited to this task because they are designed to compute the edit distance of the reference text from the hypothesis text. Our task required alignment of the text to capture the exact regions of errors – the goal of which is different slightly from the edit distance computation. For the purpose we wrote code to identify different error regions in the ASR output. Often, there is no one-to-one correspondence between an error word and a reference word. Multiple reference words can be misrecognized as a single word (substitution followed by deletions) and a single reference word can be misrecognized as multiple words (substitution followed by insertions) [22][32]. Our time-based error alignment software uses word-level timestamps to group the errors appropriately, as shown in Figure 4. The output of our processing is a list of confusion pairs for each sentence. For the example in Figure 4, the confusion pairs would be: (based, *); (send it off, son-in-law); (lead, relief); (recruiter, worker); (teams, chains).

### 5.1.2 Stimuli Selection

The alignment of the bad hypothesis output from the ASR system with the reference transcripts (in section 5.1.1 above) provided us with the list of confusion pairs, with each pair corresponding to an independent error (no overlap in the time frames) the ASR system made. We note that the reference text and the list of confusion pairs can be thought of as specifying an entire "space" of possible ASR outputs: Considering the reference text as a starting point, and considering each confusion pair as an "insert an error" operator, one can imagine an entire network of possible ASR text outputs that are possible. Each ASR output contains some subset of the errors from the list of confusion pairs.

Given this space of possible ASR outputs, our goal is to identify two output texts for each reference text, with these properties:

- The output texts should reflect reasonable performance of a commercial ASR system in noise typical of a workplace setting when the speaker is not wearing a special headset microphone; so, we wanted to identify text candidates with WER of approximately 0.25 (ranging between 20% and 30%).
- We wanted to identify one text candidate that has a low WER-to-ACE ratio and another candidate with a high WER-to-ACE ratio. We selected two candidates with identical WER: one with a high ACE score, and the other with a low ACE score.

Thus, the two text candidates identified represent two possible outputs from an ASR system. The errors that appear in the texts are realistic: They were actual errors made by an ASR system, and the overall WER error rate for the sentences is approximately 0.25. We can think of one of these text candidates as being "preferred by WER" (the one with the low WER-to-ACE ratio), and the other as being "preferred by ACE" (with the high WER-to-ACE ratio).

We wrote code to execute a search procedure through the space of possibilities to identify a pair of text candidates that fit the above criteria. We executed this code on 45 text sentences that had been extracted from the verbatim script of what the human spoke in our business meeting videos, and we thereby obtained 45 pairs of ASR text output candidates (two per sentence). Example stimuli from are available here: http://latlab.ist.rit.edu/assets2017ace

---

[3]https://sourceforge.net/projects/cmusphinx/files/Acoustic%20and%20Language%20Models/

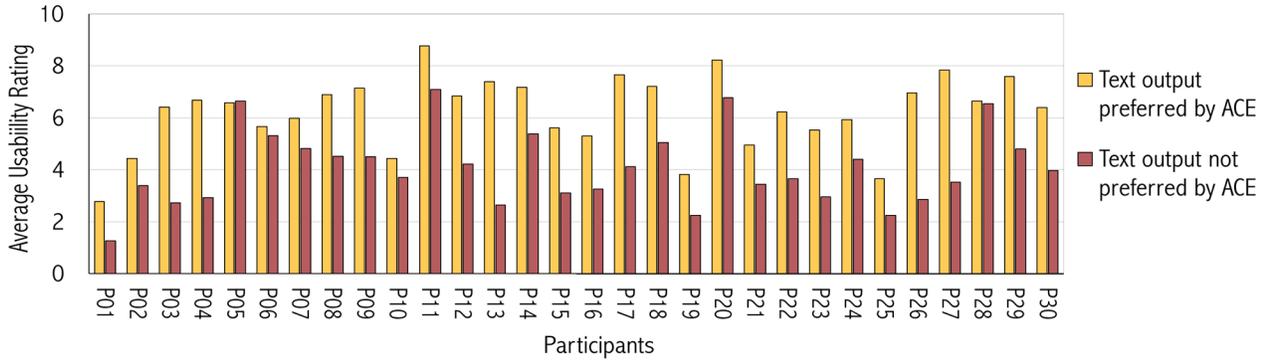

Figure 6: Average usability rating variation among participants.

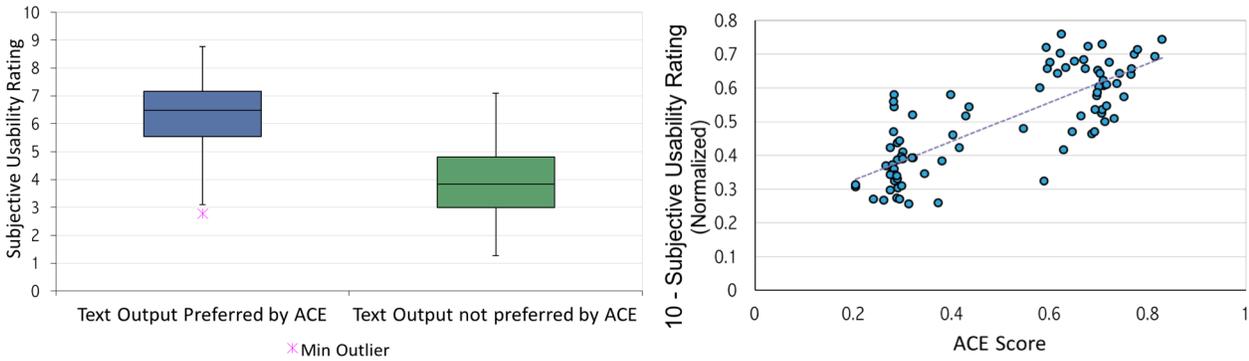

(a) Usability rating across stimuli  (b) Correlation Analysis

Figure 7: Analysis of the ACE metric with participant's usability rating.

## 5.2 Study Setup and Procedure

During the study, each participant was presented with 45 pairs of text output; each pair was displayed simultaneously, as shown in Figure 5. The reference text (what the human actually said) was provided at the top of the screen, and the two text candidates were presented on the left and right side of the screen (positioned as captioning text in a black box below the video image).

The participant was asked to provide an individual subjective quality rating for each of the two videos, using a ten-level scale (frown face to smiley face) with endpoints labeled as "Useless" and "Useful." At the beginning of the study, participants were provided with instructions on the study procedure and a practice item, prior to being presented with the 45 sentence pairs.

The WER score was identical for the two text candidates that were shown in each pair; across all 45 pairs, the WER was in the range of 0.25 to 0.3. The two versions of text differed in their ACE score; one had a higher ACE score while other had a lower score. The presentation of text candidates on the left or right side was randomized throughout the study.

## 5.3 Participants

We recruited participants from the Rochester Institute of Technology and surrounding campus community. We collected data from 30 DHH participants (age distribution with mean = 23.53 and standard deviation= 4.92), which included 17 men and 13 women. Among our participants, 14 people self-identified as deaf, 8 people identified themselves as Deaf, and 8 people of the participants as hard-of-hearing. All of the participants reported that they were familiar with the use of captioning technology, and they regularly used captioning when watching television programming.

## 6. RESULTS AND DISCUSSION

We collected 2700 responses in total from our 30 participants (subjective scores for each sentence, for 45 stimuli pair per participant). Figure 6 presents the average subjective judgment rating for each participant in the study, displaying their average score across all text candidates that they evaluated: the text output preferred by the ACE score (with high WER-to-ACE ratio) and the text output not preferred by ACE. Figure 7(a) presents the summarized response data across all participants, and Figure 7(b) visualizes a linear correlation best-fit line for the relationship between the ACE scores of each sentence and the participants' subjective judgment rating.

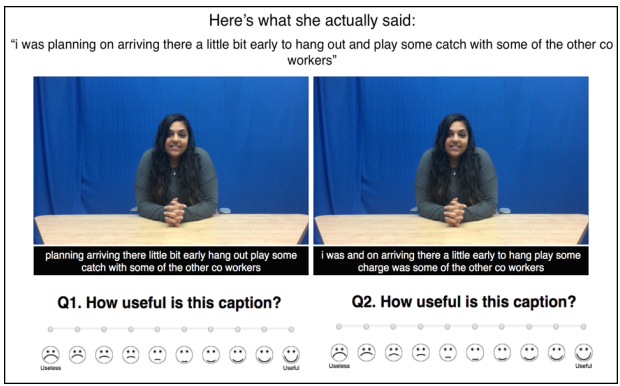

Figure 5: Screenshot from the study, with side-by-side comparison of caption-text automatically generated by ASR. Each pair of texts (left and right) have identical WER scores, but one text in each pair was preferred by our ACE metric.

Hypothesis H1 considered whether DHH users reported a subjective preference for captions that were predicted as "better" by our ACE metric, as compared to the captions that were predicted as "worse" by our ACE metric (high WER-to-ACE ratio vs. low WER-to-ACE ratio). The median difference (subjective score on "better" texts – subjective score on "worst" texts) was 2.5; the DHH users had higher subjective ratings for texts that had been preferred by the ACE metric. A boxplot summarizing the subjective rating scores from the participants for each stimulus (text predicted as "better" by ACE vs text predicted as "worse" by ACE) is shown in Figure 7(a). The distribution of the two groups differed significantly (Wilcoxon signed-rank W= 643394.00, N = 1350, N_test = 1226, P-Value < 0.0001). Thus, hypothesis H1 was supported: DHH users preferred predictions from the ACE metric.

For Hypothesis H2a, we considered if the usability scores from DHH users correlated with the ACE score significantly. We computed Spearman Rho Correlation score for the two scores. The correlation coefficient was found to be rho = 0.742791 with a p-value <0.0001. Figure 7(b) shows the corresponding correlation graph. This supports H2a: DHH user's judgments on the usability of the caption text correlated with scores from ACE metric.

For Hypothesis H2b, we performed significant difference testing on the correlations between 1) the human subjective preferences and the ACE score ($r_{ha}$) and, 2) the human subjective preferences and the WER score ($r_{hw}$). We performed a Fisher r-to-z transformation in order to perform an asymptotic z-test. For $r_{ha} = 0.742791$ and $r_{hw} = 0.108852$, we found a significant difference between the two coefficients (z-score = 5.771, 1-tail P-value < 0.0001). Thus, hypothesis H2b was supported: The subjective judgment of DHH participants about the quality of ASR captions was more highly correlated with ACE, as compared to their correlation with WER.

## 7. CONCLUSIONS AND FUTURE WORK

We have designed a new evaluation metric that analyzes the output of ASR systems to predict the impact of various ASR recognition errors on the usability of automatically generated captions for DHH users, and we have compared this new ACE metric to the traditional WER metric in a study with DHH participants. In a side-by-side comparison of pairs of ASR text output with identical WER score, the texts favored by our new metric were also preferred by DHH participants. Our metric also had significantly higher correlation with DHH participants' subjective scores on caption usability, as compared to the correlation between WER and their scores. Our new ACE metric can be used to select ASR systems for captioning applications, and researchers may utilize ACE as a loss function when training ASR systems for DHH captioning applications. The error impact module used in the metric could also find its use in error marking systems alongside word uncertainty measures to indicate readers of harmful errors in the caption.

While we have identified word predictability and semantic distance as useful predictors of the usability of an automatically generated caption text, there are still limitations in our metric, which we will address in future work, e.g., empirically evaluating alternatives to the use of *maximum* for combining error impact sub-scores (as discussed in section 3.4), e.g. to accommodate longer texts with multiple sentences. Rather than using word2vec similarity scores alone to compute semantic distance, we may also explore additional semantic features (such as sentiment). We also plan to conduct a user-study with a larger number of users to explore the relationship between additional linguistic features for identifying important words, to enable us to build and compare metrics that predict the understandability of a text; in particular, we may train a statistical model of feature weights on the usability scores collected from a larger number of DHH participants on diverse stimuli. More broadly, we will also continue our research on improving ASR based captioning for DHH users, as described in section 1.

## 8. ACKNOWLEDGMENTS

This material was based on work supported by the National Technical Institute for the Deaf (NTID). We are grateful to Christopher Caulfield, who assisted with data collection for this study, and to our collaborators Larwan Berke, Michael Stinson, Lisa Elliot, Donna Easton, and James Mallory.